\documentclass[twocolumn,twoside,5p]{elsarticle}
\usepackage[utf8]{inputenc}
\usepackage[T1]{fontenc}

\usepackage{pgfplots}
\pgfplotsset{compat=newest}

\usepackage{graphicx}
\usepackage{rotating}
\usepackage{pifont}
\usepackage{xcolor}

\usepackage{isotope}

\usepackage{scalefnt}

\usepackage{amsmath}
\usepackage{amssymb}
\usepackage{amsfonts}
\usepackage{bm}
\usepackage{xfrac}
\usepackage{braket}

\usepackage{lmodern}
\usepackage{microtype}

\usepackage{scalefnt}

\usepackage{hyperref}

\usepackage[nameinlink]{cleveref}

\pgfplotsset{compat=newest}

\definecolor{indianred}{rgb}{0.86, 0.08, 0.24}
\definecolor{royalblue}{rgb}{0.25, 0.41, 0.88}
\definecolor{darkorange}{rgb}{1.0, 0.55, 0}
\definecolor{mediumseagreen}{rgb}{0.24, 0.70, 0.44}
\definecolor{purple}{rgb}{0.5, 0, 0.5}
\definecolor{cyan3}{rgb}{0, 0.80, 0.80}

\newcommand{\symboltriangleup}[1][black]{{\color{#1}\scalefont{0.9}{\raisebox{1.5ex}{\begin{turn}{180}$\blacktriangledown$\end{turn}}}}}

\newcommand{\symbolbox}[1][black]{{\color{#1}\scalefont{0.75}$\blacksquare$}}
\newcommand{\symbolcircle}[1][black]{{\color{#1}\scalefont{0.75}\ding{108}}}

\newcommand{\symboldiamondsym}[1][black]{{\color{#1}\scalefont{0.75}\raisebox{-.2ex}{\begin{turn}{45}$\blacksquare$\end{turn}}}}

\newcommand{\symbolstar}[1][black]{{\color{#1}\raisebox{0.2ex}{$\bigstar$}}}

\newcommand{\bluecircle}{{\scalefont{0.9}\symbolcircle[royalblue]}}

\newcommand{\symbolpentagon}[1][black]{\color{#1}\scalefont{0.75}\raisebox{0.8ex}{\pgfuseplotmark{pentagon*}} } 
\newcommand{\purplepentagon}{{\scalefont{0.9}\symbolpentagon[purple]}}

\newcommand{\greentriangleup}{{\scalefont{0.9}\symboltriangleup[mediumseagreen]}}

\newcommand{\orangediamond}{{\scalefont{0.9}{\scalefont{0.8}\symboldiamondsym[darkorange]}}}

\newcommand{\redsquare}{{\scalefont{0.9}\symbolbox[indianred]}}

\newcommand{\orangestar}{{\scalefont{0.9}{\scalefont{0.8}\symbolstar[darkorange]}}}

\definecolor{plot1}{rgb}{0.86, 0.08, 0.24}
\definecolor{plot2}{rgb}{0.25, 0.41, 0.88}
\definecolor{plot3}{rgb}{1.0, 0.55, 0}
\definecolor{plot4}{RGB}{61,153,86}

\newcommand{\la}{\langle}
\newcommand{\ra}{\rangle}

\newcommand{\HFB}{\ensuremath{\vert \Phi \ra}}

\usepackage[normalem]{ulem}

\begin{document}

%authors
%\title{Bogoliubov many-body perturbation theory for open-shell nuclei}
\title{Bogoliubov Many-Body Perturbation Theory for Open-Shell Nuclei}

%\address[esnt]{ESNT, CEA, Universit\'e Paris-Saclay, 91191 Gif-sur-Yvette, France} 
\address[esnt]{ESNT, CEA-Saclay, DRF, IRFU, D\'epartement de Physique Nucl\'eaire, Universit\'e de Paris Saclay, F-91191 Gif-sur-Yvette}
\address[cea]{IRFU, CEA, Universit\'e Paris-Saclay, 91191 Gif-sur-Yvette, France} 
\address[kul]{KU Leuven, Instituut voor Kern- en Stralingsfysica, 3001 Leuven, Belgium}
%\address[dam]{CEA, DAM, DIF, F-91297 Arpajon, France}
\address[msu]{NSCL/FRIB Laboratory and Department of Physics and Astronomy, Michigan State University, East Lansing, Michigan 48824-1321, USA}
\address[tud]{Institut f\"ur Kernphysik, Technische Universit\"at Darmstadt, Schlossgartenstr. 2, 64289 Darmstadt, Germany}

\author[esnt]{A. Tichai}
\ead{alexander.tichai@cea.fr}

\author[cea]{P. Arthuis}
\ead{pierre.arthuis@cea.fr}

\author[cea,kul]{T. Duguet}
\ead{thomas.duguet@cea.fr}

\author[msu]{H. Hergert}
\ead{hergert@nscl.msu.edu}

\author[cea]{V. Som\`a}
\ead{vittorio.soma@cea.fr}

\author[tud]{R. Roth}
\ead{robert.roth@physik.tu-darmstadt.de}

\begin{abstract}
A novel Rayleigh-Schr\"odinger many-body perturbation theory (MBPT) approach is introduced by making use of a particle-number-breaking Bogoliubov reference state to tackle (near-)degenerate open-shell fermionic systems. By choosing a reference state that solves the Hartree-Fock-Bogoliubov variational problem, the approach reduces to the well-tested M{\o}ller-Plesset, i.e., Hartree-Fock based, MBPT when applied to closed-shell systems. Due to its algorithmic simplicity, the newly developed framework provides a computationally simple yet accurate alternative to state-of-the-art non-perturbative many-body approaches. At the price of working in the quasi-particle basis associated with a single-particle basis of sufficient size, the computational scaling of the method is independent of the particle number. This paper presents the first realistic applications of the method ranging from the oxygen to the nickel isotopic chains on the basis of a modern nuclear Hamiltonian derived from chiral effective field theory.
\end{abstract}


\maketitle

\paragraph*{Introduction} 
Over the past two decades the \emph{ab initio} description of nuclear structure properties has extended significantly both with respect to accessible mass numbers and to the open-shell character of the targeted system. Simplest approaches applicable to closed-shell systems start from a single-determinantal, e.g., Hartree-Fock (HF), reference state and account for \emph{dynamic correlations} via the inclusion of particle-hole excitations on top of it. In this context, a plethora of many-body frameworks have been developed to describe medium-mass systems, e.g., many-body perturbation theory (MBPT)~\cite{Ti16,Hu16,Ti17}, coupled-cluster (CC) theory~\cite{KoDe04,BaRo07,HaPa10,PiGo09,BiLa14}, self-consistent Green's functions (SCGF) theory~\cite{Dickhoff:2004xx,CiBa13,Carbone:2013eqa} or the in-medium similarity renormalization group (IMSRG) approach~\cite{TsBo11,HeBo13,Mo15,H15}. For doubly closed-shell nuclei, all of these methods agree well with quasi-exact no-core shell model (NCSM) calculations for ground-state energies of nuclei in the $A\sim 20$ regime~\cite{Hergert:2013vag}.

However, when going away from nuclear shell closures, the single-determinantal description becomes qualitatively wrong because several determinants contribute strongly to a CI expansion, requiring a proper treatment of \emph{static correlations}. 
In order to overcome this drawback, more general reference states are required, i.e., either \emph{multi-determinantal} or \emph{symmetry-broken} reference states. The latter were first used in nuclear structure through the Gorkov extension of SCGF (GSCGF) that relies on a particle-number-broken Hartree-Fock-Bogoliubov (HFB) vacuum to describe singly-open-shell nuclei~\cite{Soma:2011aj,SoCi13}. A similar extension led to designing the Bogoliubov CC formalism, although only proof-of-principle calculations limited to small model spaces and two-body forces have actually been performed so far~\cite{Si15}. In parallel, multi-determinantal reference states were successfully applied in the multi-reference extension of the IMSRG (MR-IMSRG)~\cite{Hergert:2017kx}. The first MR-IMSRG applications used particle-number-projected (PNP) HFB states~\cite{Hergert:2013vag,Hergert:2014iaa,Hergert:2017kx}. More recently, solutions of no-core shell model (NCSM) calculations~\cite{NaQu09,RoLa11,BarNa13} in a small model space were employed, leading to the so-called in-medium no-core shell model (IM-NCSM)~\cite{Geb17}, and proof-of-principle calculations with angular-momentum projected HFB states were presented in \cite{Hergert:2018wmx}. With the revival of perturbative techniques in nuclear structure theory~\cite{Ti16,Hu16} the concept of multi-determinantal reference states inspired the development of a MBPT variant based on a NCSM reference state in a small model space, yielding the perturbatively-improved no-core shell model (NCSM-PT). This method has allowed the first description of medium-mass nuclei with even and odd mass numbers on an equal footing~\cite{Ti17}. In general the use of a perturbative framework is computationally advantageous since it obviates the storage of large tensors like, e.g., cluster amplitudes in CC theory or the flowing Hamiltonian in IMSRG, and, furthermore, does not require a solution of a numerically more challenging non-linear set of equations.

Even though pioneering work based on symmetry-broken reference states was done within the GSCGF framework, similar ideas have scarcely been employed in \emph{ab initio} calculations. One reason is that symmetry breaking cannot occur in finite quantum systems, hence the explicitly broken symmetry must eventually be restored, which has been a long-standing challenge already on a formal level.
While the design of a proper symmetry-restoration protocol remains yet to be formulated within the GSCGF framework, full-fledged symmetry-broken and -restored MBPT and CC formalisms have been recently designed to consistently restore the symmetry at any truncation order~\cite{Du15,Du16}. 
The spin-projected CC version of this formalism~\cite{Du15} has been transferred and implemented successfully on the Hubbard model and on molecule dissociation~\cite{Qiu17}.

While the full details of the newly derived Bogoliubov many-body perturbation theory (BMBPT) formalism will be described in a forthcoming publication~\cite{Art18b}, its first full-fledged implementation in large model spaces with an approximate inclusion of three-body forces via normal-ordering techniques is presented in this letter. We investigate ground-state energies along complete medium-mass isotopic chains with further emphasis on two-neutron separation energies to monitor footprints of nuclear shell closures. 
Whenever possible, BMBPT calculations are benchmarked against well-established non-perturbative IT-NCSM, GSCGF, and MR-IMSRG results for the same input Hamiltonian.

\paragraph*{Many-body formalism}\label{BMBPT}
Bogoliubov MBPT is an expansion of the exact A-body ground-state energy in perturbations around a (possibly) symmetry-breaking reference state. In semi-magic nuclei, the relevant symmetry is the $U(1)$ global gauge symmetry associated with particle number conservation. Breaking $U(1)$ symmetry permits to efficiently deal with Cooper pair's instability associated with the superfluid character of open-shell nuclei. The degeneracy of a Slater determinant with respect to particle-hole excitations is lifted via the use of a Bogoliubov reference state and commuted into a degeneracy with respect to symmetry transformations of the group. 
As a consequence, the ill-defined (i.e.\@ singular) expansion of exact quantities with respect to a Slater determinant is replaced by a well-behaved one\footnote{Extending the treatment to doubly open-shell nuclei also requires a treatment of the $SU(2)$ symmetry associated with the conservation of angular momentum.}. 

Eventually, the degeneracy with respect to $U(1)$ transformations must also be lifted by restoring the symmetry. 
However, BMBPT only restores the symmetry in the limit of an all-order resummation, and, therefore retains a symmetry contamination at any finite order.  
While BMBPT is presently used as a stand-alone approach it eventually provides the first step towards the implementation of the particle-number projected BMBPT (PNP-BMBPT) which exactly restores good particle number at any truncation order~\cite{Du16}. 
While the present focus is on BMBPT, the next step will consist of implementing PNP-BMBPT.

The formalism is based on the introduction of the Bogoliubov reference state
\begin{align}
\HFB \equiv \mathcal{C} \prod_k \beta_k \vert 0 \ra\, ,
\end{align}
where $\mathcal{C}$ is a complex normalization constant and $\vert 0 \ra$ denotes the physical vacuum. The Bogoliubov state is a vacuum for the quasi-particle operators $\beta^\dagger_k, \beta_k$ that are obtained from the creation and annihilation operators of our chosen single-particle basis via the transformation
\begin{subequations}
\begin{align}
\beta_k &\equiv \sum_p U^*_{pk} c_p + V_{pk} c^\dagger_p\, , \\
\beta_k^\dagger &\equiv \sum_p U^*_{pk} c^\dagger_p + V_{pk} c_p \, .
\end{align}
\end{subequations}
While other choices are possible~\cite{Art18b}, $\HFB$ is presently obtained by solving the Hartree-Fock-Bogoliubov variational problem. The transformation matrices $(U,V)$ consist of the eigenvectors of the HFB eigenvalue equation~\cite{RiSc80}, and the quasi-particle energies $\{E_k > 0\}$ are the corresponding eigenvalues. This fixes the reference state and corresponds to the M{\o}ller-Plesset implementation of the otherwise more general Rayleigh-Schr\"odinger BMBPT formalism. 

While the HFB reference state is not an eigenstate of the particle-number operator $A$, the expectation value of $A$ is constrained to match the number of particles $A_0$ of the targeted system. It is enforced in the HFB iteration via the use of a Lagrange multiplier $\lambda$ in the minimization of the grand potential $\Omega \equiv H - \lambda A$ expectation value. In actual applications, separate Lagrange multipliers are used to constrain proton and neutron numbers separately.

In the next step, the grand potential $\Omega$ is normal ordered with respect to the HFB reference state
\begin{align}
\Omega &= \overbrace{\Omega^{00}}^{\displaystyle  \Omega^{[0]}}  + 
 \overbrace{\Omega^{20} + \Omega^{11} +\Omega^{02}}^{\displaystyle  \Omega^{[2]}}  \notag \\
&\phantom{=} +   \underbrace{\Omega^{40} + \Omega^{31} +\Omega^{22} +\Omega^{13} +\Omega^{04}}_{\displaystyle  \Omega^{[4]}} \, ,
\label{eq:NO}
\end{align} 
%%%
% Old version
%
% \begin{align}
% \Omega &= \Omega^{[0]} + \Omega^{[2]} + \Omega^{[4]} \notag \\
% &\equiv \Omega^{00} + \Big[ \Omega^{20} + \Omega^{11} +\Omega^{02} \Big ] \notag \\
% &\phantom{=} + \Big[ \Omega^{40} + \Omega^{31} +\Omega^{22} +\Omega^{13} +\Omega^{04}\Big ]\, ,
% \end{align} 
%%%
where $\Omega^{ij}$ denotes the normal-ordered component involving $i$ ($j$) quasi-particle creation (annihilation) operators, e.g., 
\begin{align}
\Omega^{31} &\equiv \frac{1}{3!}\sum_{k_1 k_2 k_3 k_4}  \Omega^{31}_{k_1 k_2 k_3 k_4}
   \beta^{\dagger}_{k_1}\beta^{\dagger}_{k_2}\beta^{\dagger}_{k_3}\beta_{k_4} \, .
\end{align} 
Thus, $\Omega^{00}$ is the expectation value of $\Omega$ in $\HFB$, $\Omega^{[2]}$ is an effective, i.e., normal-ordered, one-body operator and $\Omega^{[4]}$ is an effective two-body one. Working in the normal-ordered two-body approximation (NO2B)~\cite{Roth:2011vt} in the quasi-particle representation\footnote{We emphasize that the NO2B approximation does not break particle number itself, i.e., the truncated grand potential does commute with $A$.}, the residual three-body part $\Omega^{[6]}$ is presently discarded. Details on the normal-ordering procedure as well as expressions of the matrix elements of each operator $\Omega^{ij}$ in terms of the original matrix elements of the Hamiltonian and of the $(U,V)$ matrices can be found in Ref.~\cite{Si15}.

To set up the perturbation theory, the Hamiltonian (i.e.\@ grand potential) must be partitioned into an one-body unperturbed part $\Omega_{0}$ and a residual part $\Omega_{1}$, i.e.,
\begin{equation}
\label{split1}
\Omega = \Omega_{0} + \Omega_{1} \, .
\end{equation} 
Using a HFB reference state, $\Omega$ appearing in Eq.~\ref{eq:NO} is already naturally partitioned given that $\Omega^{20}=\Omega^{02}=0$ and that $\Omega^{11}$ is in diagonal form, i.e.,  
\begin{subequations}
\label{perturbation}
\begin{align}
\Omega_{0} &\equiv \Omega^{00} + \sum_{k} E_k \beta^{\dagger}_k \beta_k \label{perturbation1} \, , \\
\Omega_{1} &\equiv  \Omega^{40} + \Omega^{31} +\Omega^{22} +\Omega^{13} +\Omega^{04} \label{perturbation2} \, ,
\end{align}
\end{subequations}
with $E_k > 0$ for all $k$. Introducing many-body states containing even numbers of quasi-particle excitations of the vacuum
\begin{equation}
| \Phi^{k_1 k_2\ldots} \rangle \equiv \beta^{\dagger}_{k_1} \, \beta^{\dagger}_{k_2} \,  \ldots  |  \Phi \rangle \, , 
\end{equation}
the unperturbed system is fully characterized by its complete set of orthonormal eigenstates in Fock space
\begin{subequations}
\begin{align}
\Omega_{0}\, |  \Phi \rangle &= \Omega^{00} \, |  \Phi \rangle \, , \\
\Omega_{0}\, |  \Phi^{k_1 k_2\ldots} \rangle &= \left[\Omega^{00} + E_{k_1 k_2 \ldots}\right] |  \Phi^{k_1 k_2\ldots} \rangle  \label{phi} \, ,
\end{align}
\end{subequations}
where the strict positivity of unperturbed excitations $E_{k_1 k_2 \ldots} \equiv E_{k_1} + E_{k_2} +\ldots $ characterizes the lifting of the particle-hole degeneracy authorized by the spontaneous breaking of $U(1)$ symmetry in open-shell nuclei at the mean-field level.

With these ingredients at hand, the perturbation theory can be entirely worked out algebraically and/or diagrammatically. 
This can equally be done on the basis of a (imaginary) time-dependent formalism or of a time-independent formalism. While the former framework leads to working with Feynman (time-dependent) diagrams, the latter makes direct use of Goldstone (time-ordered) diagrams. The complete Rayleigh-Schr\"odinger BMBPT formalism, including the automatic generation and evaluation of all possible diagrams appearing at an arbitrary order $n$ on the basis of 2N and full 3N interactions will be presented elsewhere~\cite{Art18a,Art18b}. 

The BMBPT expression for the grand potential can be written in compact form as a Goldstone-like formula 
\begin{align}
E_0 - \lambda A_0 = \la \Phi \vert \Omega \sum_{k=0}^\infty \Big( \frac{1}{\Omega^{00}- \Omega_1} \Omega_1 \Big)^{k-1} \HFB_c \, .
\label{eq:Goldstone}
\end{align}
The lower index $c$ indicates that only connected diagrams contribute to the expansion. Thus, BMBPT is a size-extensive many-body framework that properly scales with system size. As a result of Wick's theorem, the first three orders contribute to Eq.~\ref{eq:Goldstone} according to 
\begin{align*}
E^{(1)}_0 - \lambda A_0^{(1)} &=  +\Omega^{00} \, ,  \\
E^{(2)}_0 - \lambda A_0^{(2)} &= -\frac{1}{24} \sum_{k_1 k_2 k_3 k_4} \frac{\Omega^{40}_{k_1k_2k_3k_4} \Omega^{04}_{k_1k_2 k_3 k_4} }{E_{k_1k_2k_3k_4}}  \, , \\
E^{(3)}_0 - \lambda A_0^{(3)} &= +\frac{1}{8} \sum_{\substack{k_1 k_2 k_3 \\ k_4 k_5 k_6 }} \frac{\Omega^{40}_{k_1k_2k_3k_4} \Omega^{22}_{k_5k_6 k_2 k_3} \Omega^{04}_{k_1k_5 k_6 k_4} }{E_{k_1 k_2 k_3 k_4} E_{k_1 k_5k_6k_4}}  \, .
\end{align*}
The lifting of the degeneracy with respect to particle-hole excitations in open-shell nuclei implies that the energy denominators in the perturbative corrections are non-singular and well behaved. 
Indeed, the HFB quasi-particle energies are bounded from below by the superfluid pairing gap at the Fermi energy, i.e., $\text{Min}_{k} \{E_k\} \geq \Delta_{\text{F}} > 0$. This would not be true in standard MBPT based on a Slater determinant reference state where energy denominators associated with particle-hole excitations within the open shell would be zero. Of course, BMBPT does strictly reduce to standard MBPT in a closed-shell system~\cite{Art18b}. In particular, the single third-order diagram whose algebraic expression is given above generates the three, i.e., particle-particle, hole-hole and particle-hole, third-order HF-MBPT diagrams~\cite{Art18b}. This reduction of the number of diagrams at any order $n$ is a consequence of working in a quasi-particle representation that does not distinguish particle and hole states. Conversely, all summations over quasi-particle labels run over the entire dimension of the one-body Hilbert space, which significantly increases the computational cost compared to standard MBPT.  In any case, low-order BMBPT corrections only induce low polynomial scaling with respect to quasi-particle summation and do not suffer from the storage of large tensors as more sophisticated all-order many-body approaches such as CC or IMSRG. 

As Eq.~\ref{eq:Goldstone} stipulates, the extraction of the binding energy at a given order $n$ requires the subtraction of the Lagrange term computed at the same order. Computing $A_0^{(n)}$ can be done straightforwardly by replacing the leftmost operator $\Omega$ by $A$ in Eq.~\ref{eq:Goldstone}~\cite{Art18b}. As the reference state is constrained to have the correct particle number on average, it implies that $A_0^{(1) } = A_0$. Working with a HFB reference state, it can be shown that $A_0^{(2)}=0$ due to the fact that $\Omega^{20}=\Omega^{02}=0$. Consequently, the first correction to the average particle number appears at third order such that it becomes $A_0 + A_0^{(3)} \neq A_0$, i.e., it does not match the particle number of the targeted system. This feature requires an iterative BMBPT scheme in order for the particle number to be correct at order $n \geq 3$. To do so, one needs to rerun the HFB calculation with a shifted chemical potential such that, through a series of iterations, one eventually obtains, e.g., $A_0^{(1)} + A_0^{(3)} = A_0$. Such an iterative procedure has not been implemented yet in the third-order results shown below, hence they contain an associated contamination $\Delta E^{(3)}_0$\footnote{We subsequently denote preliminary third-order BMBPT results without particle-number adjustment by BMBPT$(3^*)$ to indicate this contamination.}.

\begin{figure*}[t!]
\centering
\includegraphics[width=1.0\textwidth]{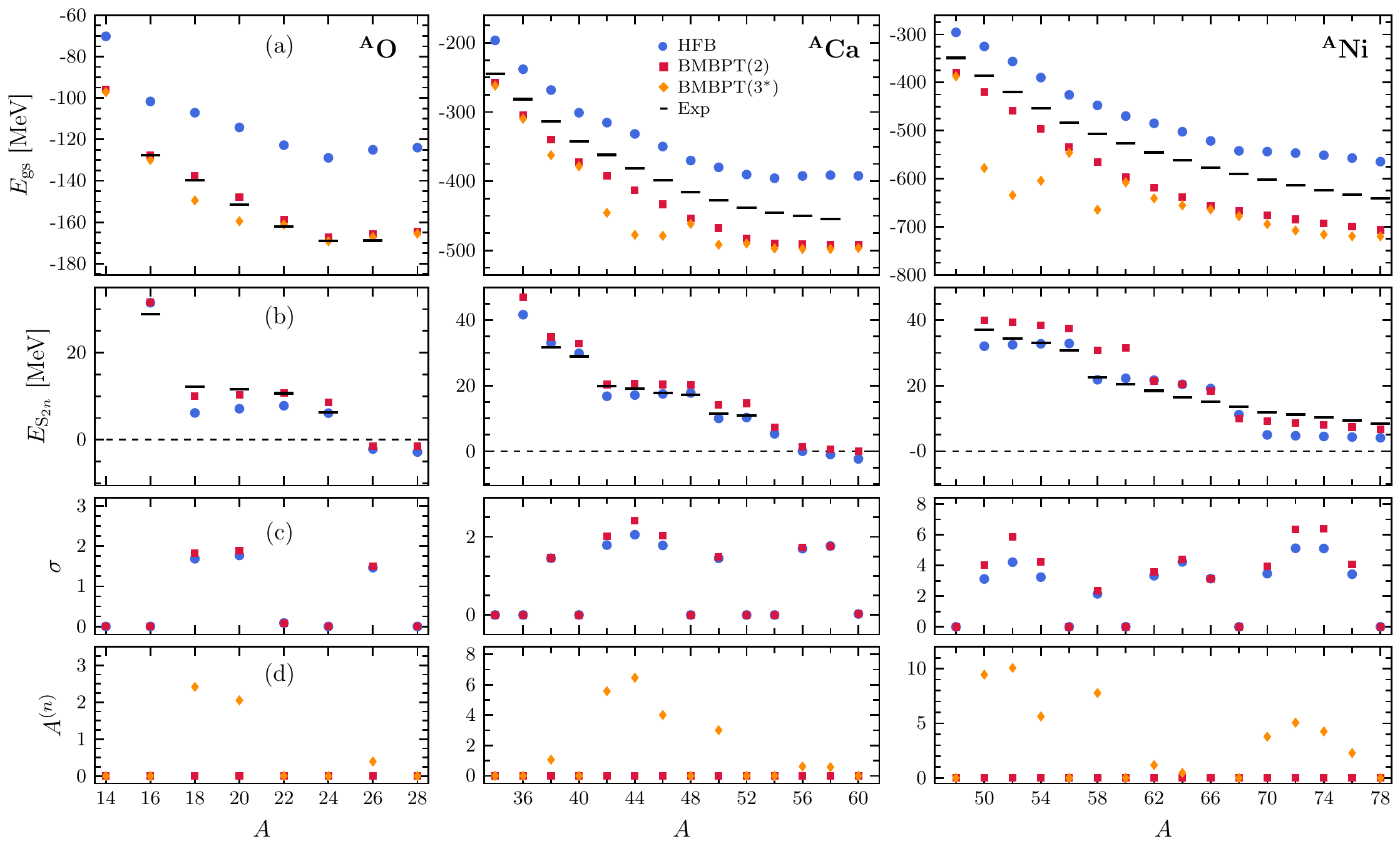}
\caption{(Color online) Systematics along O, Ca and Ni isotopic chains: (a) absolute binding energy, (b) two-neutron separation energy, (c) neutron-number dispersion, (d) perturbative correction to the average neutron number. Plot markers correspond to HFB (\bluecircle), second-order BMBPT (\redsquare) and third-order BMBPT (\orangediamond). Experimental values are shown as black bars \cite{WaAu12}.
}
\label{fig:chain}
\end{figure*}

\paragraph*{Hamiltonian}\label{Hamiltonian}
The nuclear Hamiltonian used in this work is derived from chiral effective field theory. 
It combines a chiral two-nucleon (2N) interaction at next-to-next-to-next-to leading order with a cutoff of $\Lambda_{2N}=500 \,\text{MeV}$~\cite{EnMa03} with a three-nucleon (3N) interaction\footnote{We still use the original value of $c_D$, although it was recently found that this does not reproduce the triton half-life. This interaction still provides a valuable starting point for the comparison of many-body approaches.} at next-to-next-to leading order with a local regulator based on a cutoff of $\Lambda_{3N}=400 \,\text{MeV}$~\cite{Na07,Roth:2011vt}.

The Hamiltonian is further softened using a Similarity Renormalization Group (SRG) transformation with a flow parameter $\alpha=0.08\,\text{fm}^4$~\cite{BoFu07,HeRo07,RoRe08,RoLa11,JuMa13}. This transformation induces many-nucleon forces that are included consistently up to the 3N level, i.e., chiral and induced many-body forces beyond that level are neglected. SRG-evolved Hamiltonians have already been used in a number of medium-mass calculations and have been shown to be soft enough to be used successfully in MBPT calculations~\cite{Ti16}.

\paragraph*{Low-order results in mid-mass nuclei}\label{loworder}
All calculations are performed using the eigenbasis of a spherical harmonic oscillator with frequency $\hbar \Omega=20$ MeV\footnote{The chosen value was confirmed to be close to the variational minimum from IMSRG calculations. A systematic study of variations of the oscillator frequency in the BMBPT framework is postponed to a future publication~\cite{Art18b}.}. One- and two-body operators are represented using all states up to $e_\text{max} = (2n +l ) _\text{max} =12$. Three-body matrix elements on the other hand only use a subset of the triplets built from the truncated basis such that their corresponding excitations are limited to $E_{3\text{max}}=14$. For the Hamiltonian employed here, this has proven sufficient up to heavy nickel isotopes~\cite{Binder:2014fk}.

Calculations are presently restricted to even-even semi-magic nuclear ground states characterized by  $J^\Pi=0^+$. This enables the use of angular-momentum coupling techniques to solve the HFB equations and compute the perturbative corrections. Furthermore, perturbative corrections displayed above are recast into traces over matrix products that can be evaluated economically using BLAS routines. This allows a very efficient evaluation of low-order BMBPT corrections. More details, including the $J$-scheme expressions for the normal-ordered grand potential and of the perturbative corrections will be presented in a future publication~\cite{Art18b}.

Figure~\ref{fig:chain} provides systematic results of first-, second- and (preliminary) third-order BMBPT calculations along O, Ca, and Ni isotopic chains. The top panel displaying absolute binding energies demonstrates that the bulk of dynamic correlations is obtained at second order~\cite{Ti16,Ti17}. In closed-shell, sub-closed or slightly paired open-shell nuclei, the third-order contribution is consistently suppressed compared to second order and indicates a gentle behavior of low-order BMBPT corrections. The computation of fourth-order contributions will further assess the convergence behavior of low-order BMBPT contributions based on SRG-transformed Hamiltonians in the future. 

While being informative, our preliminary third-order calculations are clearly contaminated in open-shell nuclei for which the correction to the particle number is significant, e.g., in $^{42-46}$Ca and $^{50-54}$Ni. We observe that the spurious arches in the binding energy directly reflect the behavior of $A_0^{(3)}$ displayed in panel (d) of Fig.~\ref{fig:chain}. It is consistent with the fact that the contaminating term is nothing but $\Delta E^{(3)}_0 \equiv \lambda A_0^{(3)}$, leading to an overbinding whenever $A_0^{(3)}$ leads to an excess of particles as it is systematically the case here. The contamination $\Delta E^{(3)}_0$ is presently exaggerated by the fact that the employed Hamiltonian overbinds mid-mass nuclei~\cite{Binder:2014fk}, thus making the neutron chemical potential artificially large and negative. In any case, the iterative readjustment of the average particle number at the working order $n$ will eventually eliminate the spurious arches in the binding energy.

Panel (b) of Fig.~\ref{fig:chain} displays two-neutron separation energies. While results are already qualitatively correct at first order, second-order corrections are non-negligible and tend to shrink magic gaps. The behaviour is overall very satisfactory. Panel (c) shows the neutron-number dispersion $\sigma \equiv \sqrt{\la A^2 \ra -  \la A \ra ^2}$ , which typically grows with the nuclear mass. While the dispersion is bound to go to zero in the limit of an all-order resummation, the second-order contribution does not decrease it compared to HFB. This indicates the merit of exactly restoring $U(1)$ symmetry to complement low-order dynamic correlations with non-perturbative static ones, as in projected Bogoliubov CC \cite{Du16} or MR-IMSRG \cite{Hergert:2017kx,Hergert:2018wmx}. Because the dispersion changes abruptly at (sub-)shell closures, restoring good particle number will mostly affect differential quantities, e.g.,  two-neutron separation energies, around (sub-)shell closures.

\begin{figure*}[t!]
\includegraphics[width=1.0\textwidth]{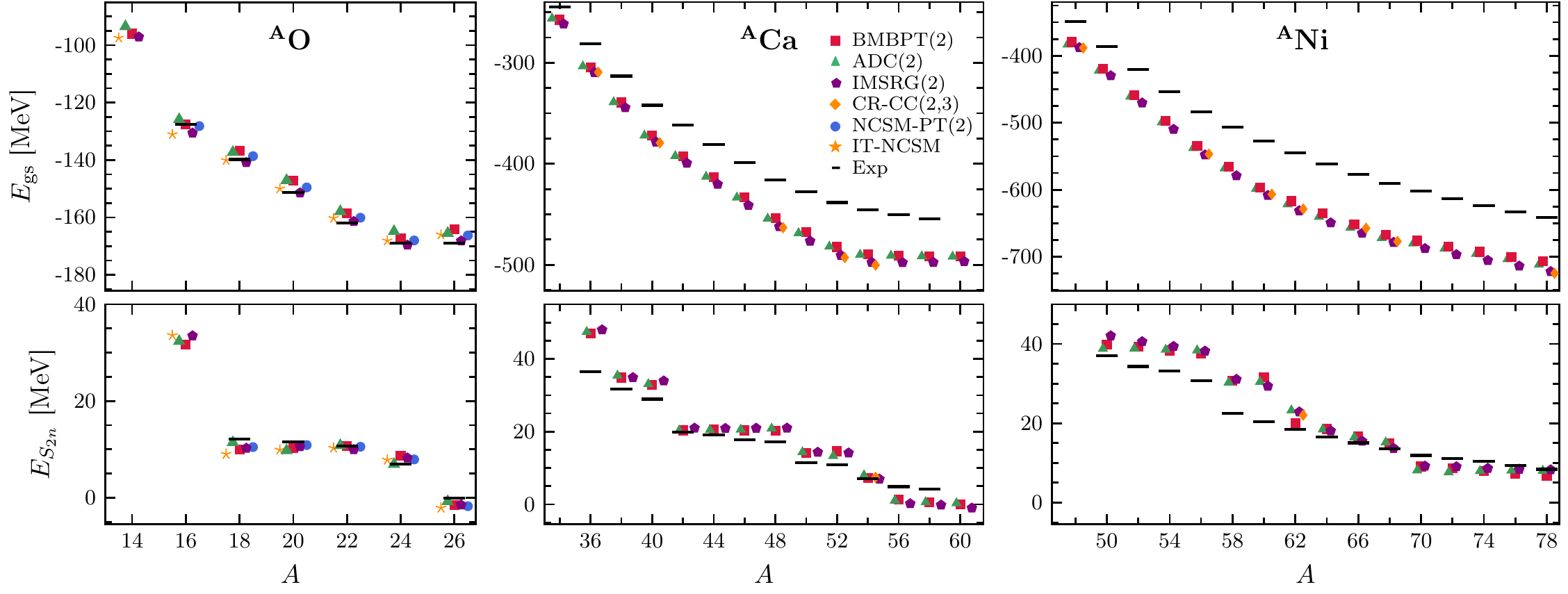}
\caption{Absolute ground-state binding energies (top) and two-neutron separation energies (bottom) along O, Ca and Ni isotopic chains. Results are displayed for second-order BMBPT (\redsquare), second-order NCSM-PT (\bluecircle), large-scale IT-NCSM (\orangestar), GSCGF-ADC(2) (\greentriangleup), MR-IMSRG(2) (\hspace{1.7pt}\purplepentagon) and CR-CC(2,3) (\orangediamond). Experimental value are shown as black bars \cite{WaAu12}. 
}
\label{fig:comp}
\end{figure*}

Figure~\ref{fig:comp} benchmarks second-order BMPBT results against well-established many-body approaches that are partially or fully non-perturbative. The Hamiltonian is the same in all calculations and numerical details associated with the basis size and the treatment of three-body forces are identical whenever possible or at least consistent. The most advanced reference, only available for O isotopes, is the importance-truncated no-core shell model (IT-NCSM) using a natural-orbital single-particle basis \cite{TiMu18}. Results from the NCSM-PT to second order are also available along the O isotopic chain \cite{Ti17}. Covering the same range of mid-mass nuclei as BMBPT, MR-IMSRG and GSCGF calculations are systematically displayed. While the IMSRG flow is truncated at the two-body level, i.e., yielding the IMSRG(2) approximation~\cite{TsBo11,H15,Hergert:2017kx}, GSCGF includes skeleton self-energy diagrams up to second order, i.e., yielding the so-called ADC(2) approximation~\cite{Schirmer83,Soma:2011aj}. Finally, closed-shell CC calculations performed at the CR-CC(2,3) level~\cite{Binder:2014fk} are added whenever available. Each of these many-body methods systematically incorporates large classes of perturbation theory diagrams beyond second-order BMBPT.

We find that second-order BMBPT ground-state energies are in very good agreement with the more sophisticated  methods for all systems under consideration, i.e., the relative deviation does not exceed $2\%$. In particular all methods are similar and in good agreement with IT-NCSM in O isotopes. MR-IMSRG(2) and NCSM-PT (when available) do provide a stronger binding compared to second-order BMBPT. On the other hand, GSCGF-ADC(2) results are very comparable to second-order BMBPT while being often slightly less bound. Of course, it will be of great interest to perform this comparison again once proper third-order and/or particle-number-restored BMBPT are systematically available. The consistency of the absolute binding energies and two-neutron separation energies provided by all the many-body methods further confirms that discrepancies with experimental data, e.g., the systematic overbinding in Ca and Ni isotopes or the incorrect behavior of $S_{2N}$ around $^{56}$Ni, reflect the shortcomings of the employed chiral Hamiltonian. CR-CC(2,3) calculations further incorporates the effect of triple excitations that are absent from MR-IMSRG(2), GSCGF-ADC(2) or second- and third-order BMBPT. Corresponding results demonstrate that a highly-accurate description of mid-mass systems requires the incorporation of triples, i.e., six-quasi-particle excitations in the language of BMBPT. The leading contributions of this type appear at fourth order in the BMBPT expansion. In addition, one should eventually consider the explicit inclusion of the 3N interaction without resorting to the NO2B approximation, as demonstrated in the CC context \cite{BiLa13,BiPi13}.

\begin{figure}[t!]
\includegraphics[width=1.\columnwidth]{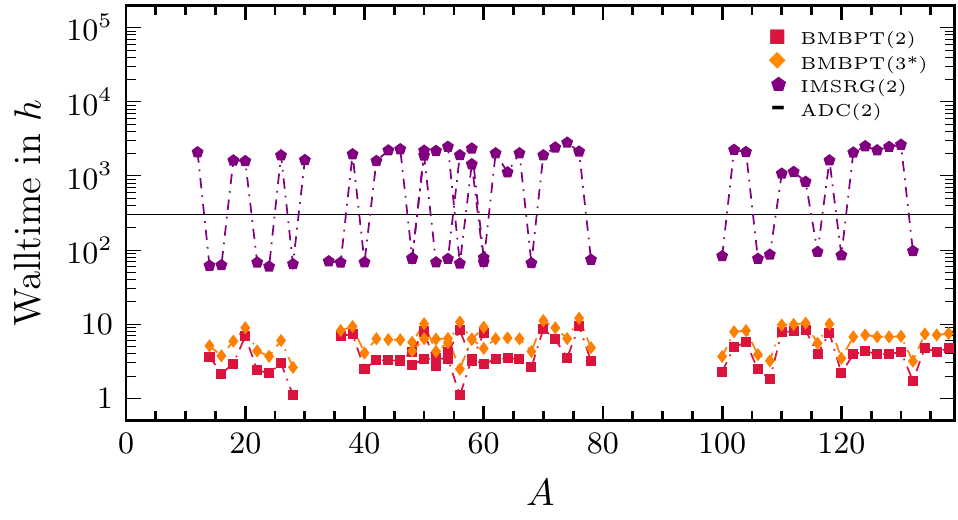}
\caption{Computational runtime versus mass number from BMBPT(2) (\redsquare), BMBPT($3^*$) (\orangediamond), MR-IMSRG(2) (\hspace{2pt}\purplepentagon) and ADC(2) calculations.
}
\label{fig:scaling}
\end{figure}

Figure~\ref{fig:scaling} provides the computational runtime in CPU hours of second- and third-order BMBPT calculations for several isotopic chains. The tin isotopic chain is included here for the record even though the corresponding results were not displayed in Figs.~\ref{fig:chain} and~\ref{fig:comp} due to the poor performance of the chiral Hamiltonian and to the lack of convergence of the calculation with respect to the $E_{3\text{max}}=14$ truncation in this mass region. BMBPT calculations were performed on an Intel Xeon X5650 computing node with 12 cores at 2.67 GHz. The runtime is essentially independent of the mass number of the system for fixed values of $e_\text{max}$ and $E_{3\text{max}}$. A typical run requires only up to $15\, \text{CPU} h$ for open-shell nuclei and as little as $6 \, \text{CPU}h$ in closed-shell nuclei. The reduction in the closed-shell case is achieved by exploiting that the Bogoliubov matrix $V$ ($U$) becomes zero for particle (hole) states when the grand potential is normal ordered, i.e., one recovers the benefit of an explicit partition between particle and hole states. In principle, we could also take advantage of the block structure of the Hamiltonian matrix with respect to isospin that disappears when normal ordering the grand potential with respect to a Bogoliubov state~\cite{Si15}. This would lead to an additional reduction by a factor of about 5, thus, making BMBPT calculations of open-shell nuclei about 10 times more expensive than genuine MBPT calculations of closed-shell systems. 

Most importantly, Fig.~\ref{fig:scaling} demonstrates that third-order BMBPT calculations generate results similar to state-of-the-art medium-mass approaches at a computational cost that is about two orders of magnitude smaller, e.g., MR-IMSRG(2) requires roughly $2000\,\text{CPU} h$ per run when applied to an open-shell system.
%\footnote{The bands in Fig.~\ref{fig:scaling} typically attempts to gauge the spread of cost due to the iterative character of a given method. While the iterative character of GSCGF-ADC(2) and MR-IMSRG(2) refers to the need to solve non-linear equations to a given accuracy, the iterative character of third-order BMBPT relates to the need to eventually adapt the average particle number. Correspondly, the lower end of the band provides the cost of an elementary iteration and the upper end corresponds to 5 times that cost, which is a somewhat realistic factor in each case. \tud{[too much detail?]}}
The computational advantage of low-order BMBPT calculations over non-perturbative approaches could make BMBPT a particularly useful tool to provide cheap systematic tests of newly generated chiral EFT Hamiltonians over a wide range of nuclei. 

\paragraph*{Conclusions}\label{conclusion}
We presented the first full-fledged \emph{ab initio} application of Bogoliubov many-body perturbation theory to finite nuclei. Expanding the exact solution around a particle-number-broken Hartree-Fock-Bogoliubov reference state, this single-reference many-body perturbation theory is systematically applicable to genuine mid- and heavy-mass open-shell nuclei. As a first proof-of-principle investigation, systematic ground-state energies along complete isotopic chains from oxygen up to tin have been computed using a standard chiral effective field theory Hamiltonian. Low-order BMBPT calculations performed on the basis of a soft interaction was found to agree at the $2\%$ level with state-of-the-art non-perturbative many-body methods at a small fraction of the computational cost. As a matter of fact, the approach is applicable beyond the tin region without becoming computationally infeasible. For now, it is the (in)accuracy of modern Hamiltonians in heavy systems and the handling of three-body matrix elements necessary to reach model-space convergence that prevent us from performing meaningful studies on nuclei far above mass number $A \approx 100$. Furthermore, the dominance of nuclear deformations for $A > 100$ requires the additional breaking of $SU(2)$ symmetry.

Our goal is to expand BMBPT in several directions in the future. The immediate next step consists of implementing the consistent adjustment of particle-number corrections at third order, which requires an iterative evaluation of the HFB equations, of the quasi-particle normal-ordering and of the perturbative corrections. A detailed investigation of this, together with a sensitivity analysis of BMBPT results with respect to model space parameters and the similarity renormalization group transformation of the Hamiltonian, will be the content of an upcoming publication. Next, the fourth-order correction will be evaluated for high-accuracy calculations and to further probe the convergence pattern of the BMBPT expansion. In that respect, it is also of interest to test Bogoliubov reference states that are \emph{not} optimized by solving the HFB equations. While the first application is limited to ground-state energies, the underlying formalism is currently being extended to other observables, e.g., charge radii, as well as to low-lying excitation energies and electromagnetic transitions. Given our capacity to automatically generate and evaluate all diagrams appearing at an arbitrary order $n$ on the basis of 2N and 3N interactions~\cite{Art18a}, it is also of interest to test the validity of the normal-ordered two-body approximation to the full 3N interaction. As a mid term goal, we plan to implement the particle-number-restoration~\cite{Du16} at second-order to investigate the impact of the symmetry contaminations on various systems/observables. In parallel, the non-perturbative Bogoliubov CC extension of BMBPT will be implemented along the line of Ref.~\cite{Du16} in order to achieve realistic applications. On the longer term, it is of interest to implement a many-body perturbation theory that consistently breaks (and restores) both $SU(2)$ and $U(1)$ symmetries to tackle doubly open-shell nuclei~\cite{Du15,Du16}.

\paragraph*{Acknowledgements}
This publication is based on work supported in part by the framework of the Espace de Structure et de r\'eactions Nucl\'eaires Th\'eorique (ESNT) at CEA, the National Science Foundation under Grant No.~PHY-1614130, the Deutsche Forschungsgemeinschaft through contract SFB 1245, the BMBF through contract 05P15RDFN1 (NuSTAR.DA), and the Helmholtz International Center for FAIR.

Calculations were performed by using HPC resources from GENCI-TGCC (Contract No. 2018-057392) as well as the Michigan State University Institute for Cyber-Enabled Research (iCER), the J\"ulich Supercomputing Center, and the Computing Center at the TU Darmstadt.

\bibliographystyle{apsrev4-1}
\bibliography{bib_nucl}

\begin{thebibliography}{45}%
\makeatletter
\providecommand \@ifxundefined [1]{%
 \@ifx{#1\undefined}
}%
\providecommand \@ifnum [1]{%
 \ifnum #1\expandafter \@firstoftwo
 \else \expandafter \@secondoftwo
 \fi
}%
\providecommand \@ifx [1]{%
 \ifx #1\expandafter \@firstoftwo
 \else \expandafter \@secondoftwo
 \fi
}%
\providecommand \natexlab [1]{#1}%
\providecommand \enquote  [1]{``#1''}%
\providecommand \bibnamefont  [1]{#1}%
\providecommand \bibfnamefont [1]{#1}%
\providecommand \citenamefont [1]{#1}%
\providecommand \href@noop [0]{\@secondoftwo}%
\providecommand \href [0]{\begingroup \@sanitize@url \@href}%
\providecommand \@href[1]{\@@startlink{#1}\@@href}%
\providecommand \@@href[1]{\endgroup#1\@@endlink}%
\providecommand \@sanitize@url [0]{\catcode `\\12\catcode `\$12\catcode
  `\&12\catcode `\#12\catcode `\^12\catcode `\_12\catcode `\%12\relax}%
\providecommand \@@startlink[1]{}%
\providecommand \@@endlink[0]{}%
\providecommand \url  [0]{\begingroup\@sanitize@url \@url }%
\providecommand \@url [1]{\endgroup\@href {#1}{\urlprefix }}%
\providecommand \urlprefix  [0]{URL }%
\providecommand \Eprint [0]{\href }%
\providecommand \doibase [0]{http://dx.doi.org/}%
\providecommand \selectlanguage [0]{\@gobble}%
\providecommand \bibinfo  [0]{\@secondoftwo}%
\providecommand \bibfield  [0]{\@secondoftwo}%
\providecommand \translation [1]{[#1]}%
\providecommand \BibitemOpen [0]{}%
\providecommand \bibitemStop [0]{}%
\providecommand \bibitemNoStop [0]{.\EOS\space}%
\providecommand \EOS [0]{\spacefactor3000\relax}%
\providecommand \BibitemShut  [1]{\csname bibitem#1\endcsname}%
\let\auto@bib@innerbib\@empty
%</preamble>
\bibitem [{\citenamefont {Tichai}\ \emph {et~al.}(2016)\citenamefont {Tichai},
  \citenamefont {Langhammer}, \citenamefont {Binder},\ and\ \citenamefont
  {Roth}}]{Ti16}%
  \BibitemOpen
  \bibfield  {author} {\bibinfo {author} {\bibfnamefont {A.}~\bibnamefont
  {Tichai}}, \bibinfo {author} {\bibfnamefont {J.}~\bibnamefont {Langhammer}},
  \bibinfo {author} {\bibfnamefont {S.}~\bibnamefont {Binder}}, \ and\ \bibinfo
  {author} {\bibfnamefont {R.}~\bibnamefont {Roth}},\ }\href {\doibase
  10.1016/j.physletb.2016.03.029} {\bibfield  {journal} {\bibinfo  {journal}
  {Physics Letters B}\ }\textbf {\bibinfo {volume} {756}},\ \bibinfo {pages}
  {283 } (\bibinfo {year} {2016})}\BibitemShut {NoStop}%
\bibitem [{\citenamefont {Hu}\ \emph {et~al.}(2016)\citenamefont {Hu},
  \citenamefont {Xu}, \citenamefont {Sun}, \citenamefont {Vary},\ and\
  \citenamefont {Li}}]{Hu16}%
  \BibitemOpen
  \bibfield  {author} {\bibinfo {author} {\bibfnamefont {B.~S.}\ \bibnamefont
  {Hu}}, \bibinfo {author} {\bibfnamefont {F.~R.}\ \bibnamefont {Xu}}, \bibinfo
  {author} {\bibfnamefont {Z.~H.}\ \bibnamefont {Sun}}, \bibinfo {author}
  {\bibfnamefont {J.~P.}\ \bibnamefont {Vary}}, \ and\ \bibinfo {author}
  {\bibfnamefont {T.}~\bibnamefont {Li}},\ }\href@noop {} {\bibfield  {journal}
  {\bibinfo  {journal} {Physical Review C}\ }\textbf {\bibinfo {volume} {94}}
  (\bibinfo {year} {2016})}\BibitemShut {NoStop}%
\bibitem [{\citenamefont {Tichai}\ \emph
  {et~al.}(2018{\natexlab{a}})\citenamefont {Tichai}, \citenamefont
  {Gebrerufael},\ and\ \citenamefont {Roth}}]{Ti17}%
  \BibitemOpen
  \bibfield  {author} {\bibinfo {author} {\bibfnamefont {A.}~\bibnamefont
  {Tichai}}, \bibinfo {author} {\bibfnamefont {E.}~\bibnamefont {Gebrerufael}},
  \ and\ \bibinfo {author} {\bibfnamefont {R.}~\bibnamefont {Roth}},\
  }\href@noop {} {\enquote {\bibinfo {title} {Open-shell nuclei from no-core
  shell model with perturbative improvement},}\ } (\bibinfo {year}
  {2018}{\natexlab{a}}),\ \bibinfo {note} {(submitted to Physics Letters
  B)}\BibitemShut {NoStop}%
\bibitem [{\citenamefont {Kowalski}\ \emph {et~al.}(2004)\citenamefont
  {Kowalski}, \citenamefont {Dean}, \citenamefont {Hjorth-Jensen},
  \citenamefont {Papenbrock},\ and\ \citenamefont {Piecuch}}]{KoDe04}%
  \BibitemOpen
  \bibfield  {author} {\bibinfo {author} {\bibfnamefont {K.}~\bibnamefont
  {Kowalski}}, \bibinfo {author} {\bibfnamefont {D.~J.}\ \bibnamefont {Dean}},
  \bibinfo {author} {\bibfnamefont {M.}~\bibnamefont {Hjorth-Jensen}}, \bibinfo
  {author} {\bibfnamefont {T.}~\bibnamefont {Papenbrock}}, \ and\ \bibinfo
  {author} {\bibfnamefont {P.}~\bibnamefont {Piecuch}},\ }\href@noop {}
  {\bibfield  {journal} {\bibinfo  {journal} {Physical Review Letters}\
  }\textbf {\bibinfo {volume} {92}},\ \bibinfo {pages} {132501} (\bibinfo
  {year} {2004})}\BibitemShut {NoStop}%
\bibitem [{\citenamefont {Bartlett}\ and\ \citenamefont
  {Musial}(2007)}]{BaRo07}%
  \BibitemOpen
  \bibfield  {author} {\bibinfo {author} {\bibfnamefont {R.~J.}\ \bibnamefont
  {Bartlett}}\ and\ \bibinfo {author} {\bibfnamefont {M.}~\bibnamefont
  {Musial}},\ }\href@noop {} {\bibfield  {journal} {\bibinfo  {journal}
  {Reviews of Modern Physics}\ }\textbf {\bibinfo {volume} {79}},\ \bibinfo
  {pages} {291} (\bibinfo {year} {2007})}\BibitemShut {NoStop}%
\bibitem [{\citenamefont {Hagen}\ \emph {et~al.}(2010)\citenamefont {Hagen},
  \citenamefont {Papenbrock}, \citenamefont {Dean},\ and\ \citenamefont
  {Hjorth-Jensen}}]{HaPa10}%
  \BibitemOpen
  \bibfield  {author} {\bibinfo {author} {\bibfnamefont {G.}~\bibnamefont
  {Hagen}}, \bibinfo {author} {\bibfnamefont {T.}~\bibnamefont {Papenbrock}},
  \bibinfo {author} {\bibfnamefont {D.~J.}\ \bibnamefont {Dean}}, \ and\
  \bibinfo {author} {\bibfnamefont {M.}~\bibnamefont {Hjorth-Jensen}},\ }\href
  {\doibase 10.1103/PhysRevC.82.034330} {\bibfield  {journal} {\bibinfo
  {journal} {Physical Review C}\ }\textbf {\bibinfo {volume} {82}},\ \bibinfo
  {pages} {034330} (\bibinfo {year} {2010})}\BibitemShut {NoStop}%
\bibitem [{\citenamefont {Piecuch}\ \emph {et~al.}(2009)\citenamefont
  {Piecuch}, \citenamefont {Gour},\ and\ \citenamefont {Wloch}}]{PiGo09}%
  \BibitemOpen
  \bibfield  {author} {\bibinfo {author} {\bibfnamefont {P.}~\bibnamefont
  {Piecuch}}, \bibinfo {author} {\bibfnamefont {J.~R.}\ \bibnamefont {Gour}}, \
  and\ \bibinfo {author} {\bibfnamefont {M.}~\bibnamefont {Wloch}},\ }\href
  {\doibase 10.1002/qua.22367} {\bibfield  {journal} {\bibinfo  {journal}
  {International Journal of Quantum Chemistry}\ }\textbf {\bibinfo {volume}
  {109}},\ \bibinfo {pages} {3268} (\bibinfo {year} {2009})}\BibitemShut
  {NoStop}%
\bibitem [{\citenamefont {Binder}\ \emph
  {et~al.}(2014{\natexlab{a}})\citenamefont {Binder}, \citenamefont
  {Langhammer}, \citenamefont {Calci},\ and\ \citenamefont {Roth}}]{BiLa14}%
  \BibitemOpen
  \bibfield  {author} {\bibinfo {author} {\bibfnamefont {S.}~\bibnamefont
  {Binder}}, \bibinfo {author} {\bibfnamefont {J.}~\bibnamefont {Langhammer}},
  \bibinfo {author} {\bibfnamefont {A.}~\bibnamefont {Calci}}, \ and\ \bibinfo
  {author} {\bibfnamefont {R.}~\bibnamefont {Roth}},\ }\href@noop {} {\bibfield
   {journal} {\bibinfo  {journal} {Physics Letters B}\ }\textbf {\bibinfo
  {volume} {736}},\ \bibinfo {pages} {119} (\bibinfo {year}
  {2014}{\natexlab{a}})}\BibitemShut {NoStop}%
\bibitem [{\citenamefont {Dickhoff}\ and\ \citenamefont
  {Barbieri}(2004)}]{Dickhoff:2004xx}%
  \BibitemOpen
  \bibfield  {author} {\bibinfo {author} {\bibfnamefont {W.~H.}\ \bibnamefont
  {Dickhoff}}\ and\ \bibinfo {author} {\bibfnamefont {C.}~\bibnamefont
  {Barbieri}},\ }\href@noop {} {\bibfield  {journal} {\bibinfo  {journal}
  {Prog. Part. Nucl. Phys.}\ }\textbf {\bibinfo {volume} {52}},\ \bibinfo
  {pages} {377} (\bibinfo {year} {2004})}\BibitemShut {NoStop}%
\bibitem [{\citenamefont {Cipollone}\ \emph {et~al.}(2013)\citenamefont
  {Cipollone}, \citenamefont {Barbieri},\ and\ \citenamefont
  {Navr{\'a}til}}]{CiBa13}%
  \BibitemOpen
  \bibfield  {author} {\bibinfo {author} {\bibfnamefont {A.}~\bibnamefont
  {Cipollone}}, \bibinfo {author} {\bibfnamefont {C.}~\bibnamefont {Barbieri}},
  \ and\ \bibinfo {author} {\bibfnamefont {P.}~\bibnamefont {Navr{\'a}til}},\
  }\href@noop {} {\bibfield  {journal} {\bibinfo  {journal} {Physical Review
  Letters}\ }\textbf {\bibinfo {volume} {111}},\ \bibinfo {pages} {062501}
  (\bibinfo {year} {2013})}\BibitemShut {NoStop}%
\bibitem [{\citenamefont {Carbone}\ \emph {et~al.}(2013)\citenamefont
  {Carbone}, \citenamefont {Cipollone}, \citenamefont {Barbieri}, \citenamefont
  {Rios},\ and\ \citenamefont {Polls}}]{Carbone:2013eqa}%
  \BibitemOpen
  \bibfield  {author} {\bibinfo {author} {\bibfnamefont {A.}~\bibnamefont
  {Carbone}}, \bibinfo {author} {\bibfnamefont {A.}~\bibnamefont {Cipollone}},
  \bibinfo {author} {\bibfnamefont {C.}~\bibnamefont {Barbieri}}, \bibinfo
  {author} {\bibfnamefont {A.}~\bibnamefont {Rios}}, \ and\ \bibinfo {author}
  {\bibfnamefont {A.}~\bibnamefont {Polls}},\ }\href@noop {} {\bibfield
  {journal} {\bibinfo  {journal} {Physical Review C}\ }\textbf {\bibinfo
  {volume} {88}},\ \bibinfo {pages} {054326} (\bibinfo {year}
  {2013})}\BibitemShut {NoStop}%
\bibitem [{\citenamefont {Tsukiyama}\ \emph {et~al.}(2011)\citenamefont
  {Tsukiyama}, \citenamefont {Bogner},\ and\ \citenamefont {Schwenk}}]{TsBo11}%
  \BibitemOpen
  \bibfield  {author} {\bibinfo {author} {\bibfnamefont {K.}~\bibnamefont
  {Tsukiyama}}, \bibinfo {author} {\bibfnamefont {S.~K.}\ \bibnamefont
  {Bogner}}, \ and\ \bibinfo {author} {\bibfnamefont {A.}~\bibnamefont
  {Schwenk}},\ }\href {\doibase 10.1103/PhysRevLett.106.222502} {\bibfield
  {journal} {\bibinfo  {journal} {Physical Review Letters}\ }\textbf {\bibinfo
  {volume} {106}},\ \bibinfo {pages} {222502} (\bibinfo {year}
  {2011})}\BibitemShut {NoStop}%
\bibitem [{\citenamefont {Hergert}\ \emph
  {et~al.}(2013{\natexlab{a}})\citenamefont {Hergert}, \citenamefont {Bogner},
  \citenamefont {Binder}, \citenamefont {Calci}, \citenamefont {Langhammer},
  \citenamefont {Roth},\ and\ \citenamefont {Schwenk}}]{HeBo13}%
  \BibitemOpen
  \bibfield  {author} {\bibinfo {author} {\bibfnamefont {H.}~\bibnamefont
  {Hergert}}, \bibinfo {author} {\bibfnamefont {S.~K.}\ \bibnamefont {Bogner}},
  \bibinfo {author} {\bibfnamefont {S.}~\bibnamefont {Binder}}, \bibinfo
  {author} {\bibfnamefont {A.}~\bibnamefont {Calci}}, \bibinfo {author}
  {\bibfnamefont {J.}~\bibnamefont {Langhammer}}, \bibinfo {author}
  {\bibfnamefont {R.}~\bibnamefont {Roth}}, \ and\ \bibinfo {author}
  {\bibfnamefont {A.}~\bibnamefont {Schwenk}},\ }\href@noop {} {\bibfield
  {journal} {\bibinfo  {journal} {Physical Review C}\ }\textbf {\bibinfo
  {volume} {87}},\ \bibinfo {pages} {034307} (\bibinfo {year}
  {2013}{\natexlab{a}})}\BibitemShut {NoStop}%
\bibitem [{\citenamefont {Morris}\ \emph {et~al.}(2015)\citenamefont {Morris},
  \citenamefont {Parzuchowski},\ and\ \citenamefont {Bogner}}]{Mo15}%
  \BibitemOpen
  \bibfield  {author} {\bibinfo {author} {\bibfnamefont {T.~D.}\ \bibnamefont
  {Morris}}, \bibinfo {author} {\bibfnamefont {N.}~\bibnamefont
  {Parzuchowski}}, \ and\ \bibinfo {author} {\bibfnamefont {S.~K.}\
  \bibnamefont {Bogner}},\ }\href {\doibase 10.1103/PhysRevC.92.034331}
  {\bibfield  {journal} {\bibinfo  {journal} {Physical Review C}\ }\textbf
  {\bibinfo {volume} {92}},\ \bibinfo {pages} {034331} (\bibinfo {year}
  {2015})}\BibitemShut {NoStop}%
\bibitem [{\citenamefont {Hergert}\ \emph {et~al.}(2016)\citenamefont
  {Hergert}, \citenamefont {Bogner}, \citenamefont {Morris}, \citenamefont
  {Schwenk},\ and\ \citenamefont {Tsukiyama}}]{H15}%
  \BibitemOpen
  \bibfield  {author} {\bibinfo {author} {\bibfnamefont {H.}~\bibnamefont
  {Hergert}}, \bibinfo {author} {\bibfnamefont {S.~K.}\ \bibnamefont {Bogner}},
  \bibinfo {author} {\bibfnamefont {T.~D.}\ \bibnamefont {Morris}}, \bibinfo
  {author} {\bibfnamefont {A.}~\bibnamefont {Schwenk}}, \ and\ \bibinfo
  {author} {\bibfnamefont {K.}~\bibnamefont {Tsukiyama}},\ }\href {\doibase
  10.1016/j.physrep.2015.12.007} {\bibfield  {journal} {\bibinfo  {journal}
  {Physics Reports}\ }\textbf {\bibinfo {volume} {621}},\ \bibinfo {pages}
  {165} (\bibinfo {year} {2016})}\BibitemShut {NoStop}%
\bibitem [{\citenamefont {Hergert}\ \emph
  {et~al.}(2013{\natexlab{b}})\citenamefont {Hergert}, \citenamefont {Binder},
  \citenamefont {Calci}, \citenamefont {Langhammer},\ and\ \citenamefont
  {Roth}}]{Hergert:2013vag}%
  \BibitemOpen
  \bibfield  {author} {\bibinfo {author} {\bibfnamefont {H.}~\bibnamefont
  {Hergert}}, \bibinfo {author} {\bibfnamefont {S.}~\bibnamefont {Binder}},
  \bibinfo {author} {\bibfnamefont {A.}~\bibnamefont {Calci}}, \bibinfo
  {author} {\bibfnamefont {J.}~\bibnamefont {Langhammer}}, \ and\ \bibinfo
  {author} {\bibfnamefont {R.}~\bibnamefont {Roth}},\ }\href@noop {} {\bibfield
   {journal} {\bibinfo  {journal} {Physical Review Letters}\ }\textbf {\bibinfo
  {volume} {110}},\ \bibinfo {pages} {242501} (\bibinfo {year}
  {2013}{\natexlab{b}})}\BibitemShut {NoStop}%
\bibitem [{\citenamefont {Som{\`a}}\ \emph {et~al.}(2011)\citenamefont
  {Som{\`a}}, \citenamefont {Duguet},\ and\ \citenamefont
  {Barbieri}}]{Soma:2011aj}%
  \BibitemOpen
  \bibfield  {author} {\bibinfo {author} {\bibfnamefont {V.}~\bibnamefont
  {Som{\`a}}}, \bibinfo {author} {\bibfnamefont {T.}~\bibnamefont {Duguet}}, \
  and\ \bibinfo {author} {\bibfnamefont {C.}~\bibnamefont {Barbieri}},\
  }\href@noop {} {\bibfield  {journal} {\bibinfo  {journal} {Physical Review
  C}\ }\textbf {\bibinfo {volume} {84}},\ \bibinfo {pages} {064317} (\bibinfo
  {year} {2011})}\BibitemShut {NoStop}%
\bibitem [{\citenamefont {Som{\`a}}\ \emph {et~al.}(2014)\citenamefont
  {Som{\`a}}, \citenamefont {Cipollone}, \citenamefont {Barbieri},
  \citenamefont {Navr{\'a}til},\ and\ \citenamefont {Duguet}}]{SoCi13}%
  \BibitemOpen
  \bibfield  {author} {\bibinfo {author} {\bibfnamefont {V.}~\bibnamefont
  {Som{\`a}}}, \bibinfo {author} {\bibfnamefont {A.}~\bibnamefont {Cipollone}},
  \bibinfo {author} {\bibfnamefont {C.}~\bibnamefont {Barbieri}}, \bibinfo
  {author} {\bibfnamefont {P.}~\bibnamefont {Navr{\'a}til}}, \ and\ \bibinfo
  {author} {\bibfnamefont {T.}~\bibnamefont {Duguet}},\ }\href@noop {}
  {\bibfield  {journal} {\bibinfo  {journal} {Physical Review C}\ }\textbf
  {\bibinfo {volume} {89}},\ \bibinfo {pages} {061301} (\bibinfo {year}
  {2014})}\BibitemShut {NoStop}%
\bibitem [{\citenamefont {Signoracci}\ \emph {et~al.}(2015)\citenamefont
  {Signoracci}, \citenamefont {Duguet}, \citenamefont {Hagen},\ and\
  \citenamefont {Jansen}}]{Si15}%
  \BibitemOpen
  \bibfield  {author} {\bibinfo {author} {\bibfnamefont {A.}~\bibnamefont
  {Signoracci}}, \bibinfo {author} {\bibfnamefont {T.}~\bibnamefont {Duguet}},
  \bibinfo {author} {\bibfnamefont {G.}~\bibnamefont {Hagen}}, \ and\ \bibinfo
  {author} {\bibfnamefont {G.~R.}\ \bibnamefont {Jansen}},\ }\href {\doibase
  10.1103/PhysRevC.91.064320} {\bibfield  {journal} {\bibinfo  {journal}
  {Physical Review C}\ }\textbf {\bibinfo {volume} {91}},\ \bibinfo {pages}
  {064320} (\bibinfo {year} {2015})}\BibitemShut {NoStop}%
\bibitem [{\citenamefont {Hergert}(2017)}]{Hergert:2017kx}%
  \BibitemOpen
  \bibfield  {author} {\bibinfo {author} {\bibfnamefont {H.}~\bibnamefont
  {Hergert}},\ }\href@noop {} {\bibfield  {journal} {\bibinfo  {journal} {Phys.
  Scripta}\ }\textbf {\bibinfo {volume} {92}},\ \bibinfo {pages} {023002}
  (\bibinfo {year} {2017})}\BibitemShut {NoStop}%
\bibitem [{\citenamefont {Hergert}\ \emph {et~al.}(2014)\citenamefont
  {Hergert}, \citenamefont {Bogner}, \citenamefont {Morris}, \citenamefont
  {Binder}, \citenamefont {Calci}, \citenamefont {Langhammer},\ and\
  \citenamefont {Roth}}]{Hergert:2014iaa}%
  \BibitemOpen
  \bibfield  {author} {\bibinfo {author} {\bibfnamefont {H.}~\bibnamefont
  {Hergert}}, \bibinfo {author} {\bibfnamefont {S.~K.}\ \bibnamefont {Bogner}},
  \bibinfo {author} {\bibfnamefont {T.~D.}\ \bibnamefont {Morris}}, \bibinfo
  {author} {\bibfnamefont {S.}~\bibnamefont {Binder}}, \bibinfo {author}
  {\bibfnamefont {A.}~\bibnamefont {Calci}}, \bibinfo {author} {\bibfnamefont
  {J.}~\bibnamefont {Langhammer}}, \ and\ \bibinfo {author} {\bibfnamefont
  {R.}~\bibnamefont {Roth}},\ }\href@noop {} {\bibfield  {journal} {\bibinfo
  {journal} {Physical Review C}\ }\textbf {\bibinfo {volume} {90}},\ \bibinfo
  {pages} {041302} (\bibinfo {year} {2014})}\BibitemShut {NoStop}%
\bibitem [{\citenamefont {Navr\'atil}\ \emph {et~al.}(2009)\citenamefont
  {Navr\'atil}, \citenamefont {Quaglioni}, \citenamefont {Stetcu},\ and\
  \citenamefont {Barrett}}]{NaQu09}%
  \BibitemOpen
  \bibfield  {author} {\bibinfo {author} {\bibfnamefont {P.}~\bibnamefont
  {Navr\'atil}}, \bibinfo {author} {\bibfnamefont {S.}~\bibnamefont
  {Quaglioni}}, \bibinfo {author} {\bibfnamefont {I.}~\bibnamefont {Stetcu}}, \
  and\ \bibinfo {author} {\bibfnamefont {B.}~\bibnamefont {Barrett}},\ }\href
  {\doibase 10.1088/0954-3899/36/8/083101} {\bibfield  {journal} {\bibinfo
  {journal} {Journal of Physics G: Nuclear and Particle Physics}\ }\textbf
  {\bibinfo {volume} {36}},\ \bibinfo {pages} {083101} (\bibinfo {year}
  {2009})}\BibitemShut {NoStop}%
\bibitem [{\citenamefont {Roth}\ \emph {et~al.}(2011)\citenamefont {Roth},
  \citenamefont {Langhammer}, \citenamefont {Calci}, \citenamefont {Binder},\
  and\ \citenamefont {Navr{\'a}til}}]{RoLa11}%
  \BibitemOpen
  \bibfield  {author} {\bibinfo {author} {\bibfnamefont {R.}~\bibnamefont
  {Roth}}, \bibinfo {author} {\bibfnamefont {J.}~\bibnamefont {Langhammer}},
  \bibinfo {author} {\bibfnamefont {A.}~\bibnamefont {Calci}}, \bibinfo
  {author} {\bibfnamefont {S.}~\bibnamefont {Binder}}, \ and\ \bibinfo {author}
  {\bibfnamefont {P.}~\bibnamefont {Navr{\'a}til}},\ }\href {\doibase
  10.1103/PhysRevLett.107.072501} {\bibfield  {journal} {\bibinfo  {journal}
  {Physical Review Letters}\ }\textbf {\bibinfo {volume} {107}},\ \bibinfo
  {pages} {072501} (\bibinfo {year} {2011})}\BibitemShut {NoStop}%
\bibitem [{\citenamefont {Barrett}\ \emph {et~al.}(2013)\citenamefont
  {Barrett}, \citenamefont {Navr{\'a}til},\ and\ \citenamefont
  {Vary}}]{BarNa13}%
  \BibitemOpen
  \bibfield  {author} {\bibinfo {author} {\bibfnamefont {B.~R.}\ \bibnamefont
  {Barrett}}, \bibinfo {author} {\bibfnamefont {P.}~\bibnamefont
  {Navr{\'a}til}}, \ and\ \bibinfo {author} {\bibfnamefont {J.~P.}\
  \bibnamefont {Vary}},\ }\href {\doibase 10.1016/j.ppnp.2012.10.003}
  {\bibfield  {journal} {\bibinfo  {journal} {Progress in Particle and Nuclear
  Physics}\ }\textbf {\bibinfo {volume} {69}},\ \bibinfo {pages} {131}
  (\bibinfo {year} {2013})}\BibitemShut {NoStop}%
\bibitem [{\citenamefont {Gebrerufael}\ \emph {et~al.}(2017)\citenamefont
  {Gebrerufael}, \citenamefont {Vobig}, \citenamefont {Hergert},\ and\
  \citenamefont {Roth}}]{Geb17}%
  \BibitemOpen
  \bibfield  {author} {\bibinfo {author} {\bibfnamefont {E.}~\bibnamefont
  {Gebrerufael}}, \bibinfo {author} {\bibfnamefont {K.}~\bibnamefont {Vobig}},
  \bibinfo {author} {\bibfnamefont {H.}~\bibnamefont {Hergert}}, \ and\
  \bibinfo {author} {\bibfnamefont {R.}~\bibnamefont {Roth}},\ }\href {\doibase
  10.1103/PhysRevLett.118.152503} {\bibfield  {journal} {\bibinfo  {journal}
  {Physical Review Letters}\ }\textbf {\bibinfo {volume} {118}},\ \bibinfo
  {pages} {152503} (\bibinfo {year} {2017})}\BibitemShut {NoStop}%
\bibitem [{\citenamefont {Hergert}\ \emph {et~al.}(2018)\citenamefont
  {Hergert}, \citenamefont {Yao}, \citenamefont {Morris}, \citenamefont
  {Parzuchowski}, \citenamefont {Bogner},\ and\ \citenamefont
  {Engel}}]{Hergert:2018wmx}%
  \BibitemOpen
  \bibfield  {author} {\bibinfo {author} {\bibfnamefont {H.}~\bibnamefont
  {Hergert}}, \bibinfo {author} {\bibfnamefont {J.}~\bibnamefont {Yao}},
  \bibinfo {author} {\bibfnamefont {T.~D.}\ \bibnamefont {Morris}}, \bibinfo
  {author} {\bibfnamefont {N.~M.}\ \bibnamefont {Parzuchowski}}, \bibinfo
  {author} {\bibfnamefont {S.~K.}\ \bibnamefont {Bogner}}, \ and\ \bibinfo
  {author} {\bibfnamefont {J.}~\bibnamefont {Engel}},\ }in\ \href@noop {}
  {\emph {\bibinfo {booktitle} {{19th International Conference on Recent
  Progress in Many-Body Theories (RPMBT19) Pohang, Korea, June 25-30, 2017}}}}\
  (\bibinfo {year} {2018})\ \Eprint {http://arxiv.org/abs/1805.09221}
  {arXiv:1805.09221 [nucl-th]} \BibitemShut {NoStop}%
%%CITATION = ARXIV:1805.09221;%%
\bibitem [{\citenamefont {Duguet}(2015)}]{Du15}%
  \BibitemOpen
  \bibfield  {author} {\bibinfo {author} {\bibfnamefont {T.}~\bibnamefont
  {Duguet}},\ }\href {\doibase 10.1088/0954-3899/42/2/025107} {\bibfield
  {journal} {\bibinfo  {journal} {Journal of Physics G: Nuclear and Particle
  Physics}\ }\textbf {\bibinfo {volume} {42}},\ \bibinfo {pages} {025107}
  (\bibinfo {year} {2015})}\BibitemShut {NoStop}%
\bibitem [{\citenamefont {Duguet}\ and\ \citenamefont
  {Signoracci}(2016)}]{Du16}%
  \BibitemOpen
  \bibfield  {author} {\bibinfo {author} {\bibfnamefont {T.}~\bibnamefont
  {Duguet}}\ and\ \bibinfo {author} {\bibfnamefont {A.}~\bibnamefont
  {Signoracci}},\ }\href@noop {} {\bibfield  {journal} {\bibinfo  {journal}
  {Journal of Physics G: Nuclear and Particle Physics}\ }\textbf {\bibinfo
  {volume} {44}} (\bibinfo {year} {2016})}\BibitemShut {NoStop}%
\bibitem [{\citenamefont {Qiu}\ \emph {et~al.}(2017)\citenamefont {Qiu},
  \citenamefont {Henderson}, \citenamefont {Zhao},\ and\ \citenamefont
  {Scuseria}}]{Qiu17}%
  \BibitemOpen
  \bibfield  {author} {\bibinfo {author} {\bibfnamefont {Y.}~\bibnamefont
  {Qiu}}, \bibinfo {author} {\bibfnamefont {T.~M.}\ \bibnamefont {Henderson}},
  \bibinfo {author} {\bibfnamefont {J.}~\bibnamefont {Zhao}}, \ and\ \bibinfo
  {author} {\bibfnamefont {G.~E.}\ \bibnamefont {Scuseria}},\ }\href@noop {}
  {\bibfield  {journal} {\bibinfo  {journal} {The Journal of Chemical Physics}\
  }\textbf {\bibinfo {volume} {147}} (\bibinfo {year} {2017})}\BibitemShut
  {NoStop}%
\bibitem [{\citenamefont {Arthuis}\ \emph
  {et~al.}(2018{\natexlab{a}})\citenamefont {Arthuis}, \citenamefont {Tichai},\
  and\ \citenamefont {Duguet}}]{Art18b}%
  \BibitemOpen
  \bibfield  {author} {\bibinfo {author} {\bibfnamefont {P.}~\bibnamefont
  {Arthuis}}, \bibinfo {author} {\bibfnamefont {A.}~\bibnamefont {Tichai}}, \
  and\ \bibinfo {author} {\bibfnamefont {T.}~\bibnamefont {Duguet}},\
  }\href@noop {} {\enquote {\bibinfo {title} {Bogoliubov many-body perturbation
  theory formalism},}\ } (\bibinfo {year} {2018}{\natexlab{a}}),\ \bibinfo
  {note} {unpublished}\BibitemShut {NoStop}%
\bibitem [{\citenamefont {Ring}\ and\ \citenamefont {Schuck}(1980)}]{RiSc80}%
  \BibitemOpen
  \bibfield  {author} {\bibinfo {author} {\bibfnamefont {P.}~\bibnamefont
  {Ring}}\ and\ \bibinfo {author} {\bibfnamefont {P.}~\bibnamefont {Schuck}},\
  }\href@noop {} {\emph {\bibinfo {title} {The Nuclear Many-Body Problem}}}\
  (\bibinfo  {publisher} {Springer Verlag, New York},\ \bibinfo {year}
  {1980})\BibitemShut {NoStop}%
\bibitem [{\citenamefont {Roth}\ \emph {et~al.}(2012)\citenamefont {Roth},
  \citenamefont {Binder}, \citenamefont {Vobig}, \citenamefont {Calci},
  \citenamefont {Langhammer},\ and\ \citenamefont {Navratil}}]{Roth:2011vt}%
  \BibitemOpen
  \bibfield  {author} {\bibinfo {author} {\bibfnamefont {R.}~\bibnamefont
  {Roth}}, \bibinfo {author} {\bibfnamefont {S.}~\bibnamefont {Binder}},
  \bibinfo {author} {\bibfnamefont {K.}~\bibnamefont {Vobig}}, \bibinfo
  {author} {\bibfnamefont {A.}~\bibnamefont {Calci}}, \bibinfo {author}
  {\bibfnamefont {J.}~\bibnamefont {Langhammer}}, \ and\ \bibinfo {author}
  {\bibfnamefont {P.}~\bibnamefont {Navratil}},\ }\href {\doibase
  10.1103/PhysRevLett.109.052501} {\bibfield  {journal} {\bibinfo  {journal}
  {Physical Review Letters}\ }\textbf {\bibinfo {volume} {109}},\ \bibinfo
  {pages} {052501} (\bibinfo {year} {2012})}\BibitemShut {NoStop}%
\bibitem [{\citenamefont {Arthuis}\ \emph
  {et~al.}(2018{\natexlab{b}})\citenamefont {Arthuis}, \citenamefont {Duguet},
  \citenamefont {Tichai}, \citenamefont {Lasseri},\ and\ \citenamefont
  {Ebran}}]{Art18a}%
  \BibitemOpen
  \bibfield  {author} {\bibinfo {author} {\bibfnamefont {P.}~\bibnamefont
  {Arthuis}}, \bibinfo {author} {\bibfnamefont {T.}~\bibnamefont {Duguet}},
  \bibinfo {author} {\bibfnamefont {A.}~\bibnamefont {Tichai}}, \bibinfo
  {author} {\bibfnamefont {R.-D.}\ \bibnamefont {Lasseri}}, \ and\ \bibinfo
  {author} {\bibfnamefont {J.-P.}\ \bibnamefont {Ebran}},\ }\href@noop {}
  {\enquote {\bibinfo {title} {Automated generation and evaluation of many-body
  diagrams. {T}he program {ADG} (v1.00) {B}ogoliubov many-body perturbation
  theory.}}\ } (\bibinfo {year} {2018}{\natexlab{b}}),\ \bibinfo {note}
  {unpublished}\BibitemShut {NoStop}%
\bibitem [{\citenamefont {Wang}\ \emph {et~al.}(2012)\citenamefont {Wang},
  \citenamefont {Audi}, \citenamefont {Wapstra}, \citenamefont {Kondev},
  \citenamefont {MacCormick}, \citenamefont {Xu},\ and\ \citenamefont
  {Pfeiffer}}]{WaAu12}%
  \BibitemOpen
  \bibfield  {author} {\bibinfo {author} {\bibfnamefont {M.}~\bibnamefont
  {Wang}}, \bibinfo {author} {\bibfnamefont {G.}~\bibnamefont {Audi}}, \bibinfo
  {author} {\bibfnamefont {A.~H.}\ \bibnamefont {Wapstra}}, \bibinfo {author}
  {\bibfnamefont {F.~G.}\ \bibnamefont {Kondev}}, \bibinfo {author}
  {\bibfnamefont {M.}~\bibnamefont {MacCormick}}, \bibinfo {author}
  {\bibfnamefont {X.}~\bibnamefont {Xu}}, \ and\ \bibinfo {author}
  {\bibfnamefont {B.}~\bibnamefont {Pfeiffer}},\ }\href@noop {} {\bibfield
  {journal} {\bibinfo  {journal} {Chinese Physics C}\ }\textbf {\bibinfo
  {volume} {36}},\ \bibinfo {pages} {1603} (\bibinfo {year}
  {2012})}\BibitemShut {NoStop}%
\bibitem [{\citenamefont {Entem}\ and\ \citenamefont
  {Machleidt}(2003)}]{EnMa03}%
  \BibitemOpen
  \bibfield  {author} {\bibinfo {author} {\bibfnamefont {D.~R.}\ \bibnamefont
  {Entem}}\ and\ \bibinfo {author} {\bibfnamefont {R.}~\bibnamefont
  {Machleidt}},\ }\href@noop {} {\bibfield  {journal} {\bibinfo  {journal}
  {Physical Review C}\ }\textbf {\bibinfo {volume} {68}},\ \bibinfo {pages}
  {041001(R)} (\bibinfo {year} {2003})}\BibitemShut {NoStop}%
\bibitem [{\citenamefont {Navr\'atil}(2007)}]{Na07}%
  \BibitemOpen
  \bibfield  {author} {\bibinfo {author} {\bibfnamefont {P.}~\bibnamefont
  {Navr\'atil}},\ }\href {\doibase 10.1007/s00601-007-0193-3} {\bibfield
  {journal} {\bibinfo  {journal} {Few Body Systems}\ }\textbf {\bibinfo
  {volume} {41}},\ \bibinfo {pages} {117} (\bibinfo {year} {2007})}\BibitemShut
  {NoStop}%
\bibitem [{\citenamefont {Bogner}\ \emph {et~al.}(2007)\citenamefont {Bogner},
  \citenamefont {Furnstahl},\ and\ \citenamefont {Perry}}]{BoFu07}%
  \BibitemOpen
  \bibfield  {author} {\bibinfo {author} {\bibfnamefont {S.~K.}\ \bibnamefont
  {Bogner}}, \bibinfo {author} {\bibfnamefont {R.~J.}\ \bibnamefont
  {Furnstahl}}, \ and\ \bibinfo {author} {\bibfnamefont {R.~J.}\ \bibnamefont
  {Perry}},\ }\href@noop {} {\bibfield  {journal} {\bibinfo  {journal}
  {Physical Review C}\ }\textbf {\bibinfo {volume} {75}},\ \bibinfo {pages}
  {061001(R)} (\bibinfo {year} {2007})}\BibitemShut {NoStop}%
\bibitem [{\citenamefont {Hergert}\ and\ \citenamefont {Roth}(2007)}]{HeRo07}%
  \BibitemOpen
  \bibfield  {author} {\bibinfo {author} {\bibfnamefont {H.}~\bibnamefont
  {Hergert}}\ and\ \bibinfo {author} {\bibfnamefont {R.}~\bibnamefont {Roth}},\
  }\href@noop {} {\bibfield  {journal} {\bibinfo  {journal} {Physical Review
  C}\ }\textbf {\bibinfo {volume} {75}},\ \bibinfo {pages} {051001(R)}
  (\bibinfo {year} {2007})}\BibitemShut {NoStop}%
\bibitem [{\citenamefont {Roth}\ \emph {et~al.}(2008)\citenamefont {Roth},
  \citenamefont {Reinhardt},\ and\ \citenamefont {Hergert}}]{RoRe08}%
  \BibitemOpen
  \bibfield  {author} {\bibinfo {author} {\bibfnamefont {R.}~\bibnamefont
  {Roth}}, \bibinfo {author} {\bibfnamefont {S.}~\bibnamefont {Reinhardt}}, \
  and\ \bibinfo {author} {\bibfnamefont {H.}~\bibnamefont {Hergert}},\
  }\href@noop {} {\bibfield  {journal} {\bibinfo  {journal} {Physical Review
  C}\ }\textbf {\bibinfo {volume} {77}},\ \bibinfo {pages} {064003} (\bibinfo
  {year} {2008})}\BibitemShut {NoStop}%
\bibitem [{\citenamefont {Jurgenson}\ \emph {et~al.}(2013)\citenamefont
  {Jurgenson}, \citenamefont {Maris}, \citenamefont {Furnstahl}, \citenamefont
  {Navr{\'a}til}, \citenamefont {Ormand},\ and\ \citenamefont {Vary}}]{JuMa13}%
  \BibitemOpen
  \bibfield  {author} {\bibinfo {author} {\bibfnamefont {E.~D.}\ \bibnamefont
  {Jurgenson}}, \bibinfo {author} {\bibfnamefont {P.}~\bibnamefont {Maris}},
  \bibinfo {author} {\bibfnamefont {R.~J.}\ \bibnamefont {Furnstahl}}, \bibinfo
  {author} {\bibfnamefont {P.}~\bibnamefont {Navr{\'a}til}}, \bibinfo {author}
  {\bibfnamefont {W.~E.}\ \bibnamefont {Ormand}}, \ and\ \bibinfo {author}
  {\bibfnamefont {J.~P.}\ \bibnamefont {Vary}},\ }\href@noop {} {\bibfield
  {journal} {\bibinfo  {journal} {Physical Review C}\ }\textbf {\bibinfo
  {volume} {87}},\ \bibinfo {pages} {054312} (\bibinfo {year}
  {2013})}\BibitemShut {NoStop}%
\bibitem [{\citenamefont {Binder}\ \emph
  {et~al.}(2014{\natexlab{b}})\citenamefont {Binder}, \citenamefont
  {Langhammer}, \citenamefont {Calci},\ and\ \citenamefont
  {Roth}}]{Binder:2014fk}%
  \BibitemOpen
  \bibfield  {author} {\bibinfo {author} {\bibfnamefont {S.}~\bibnamefont
  {Binder}}, \bibinfo {author} {\bibfnamefont {J.}~\bibnamefont {Langhammer}},
  \bibinfo {author} {\bibfnamefont {A.}~\bibnamefont {Calci}}, \ and\ \bibinfo
  {author} {\bibfnamefont {R.}~\bibnamefont {Roth}},\ }\href@noop {} {\bibfield
   {journal} {\bibinfo  {journal} {Phys. Lett. B}\ }\textbf {\bibinfo {volume}
  {736}},\ \bibinfo {pages} {119 } (\bibinfo {year}
  {2014}{\natexlab{b}})}\BibitemShut {NoStop}%
\bibitem [{\citenamefont {Tichai}\ \emph
  {et~al.}(2018{\natexlab{b}})\citenamefont {Tichai}, \citenamefont {M\"uller},
  \citenamefont {Vobig},\ and\ \citenamefont {Roth}}]{TiMu18}%
  \BibitemOpen
  \bibfield  {author} {\bibinfo {author} {\bibfnamefont {A.}~\bibnamefont
  {Tichai}}, \bibinfo {author} {\bibfnamefont {J.}~\bibnamefont {M\"uller}},
  \bibinfo {author} {\bibfnamefont {K.}~\bibnamefont {Vobig}}, \ and\ \bibinfo
  {author} {\bibfnamefont {R.}~\bibnamefont {Roth}},\ }\href@noop {} {\enquote
  {\bibinfo {title} {Natural orbitals for \emph{ab initio} no-core shell-model
  calculations},}\ } (\bibinfo {year} {2018}{\natexlab{b}}),\ \bibinfo {note}
  {(unpublished)}\BibitemShut {NoStop}%
\bibitem [{\citenamefont {Schirmer}\ \emph {et~al.}(1983)\citenamefont
  {Schirmer}, \citenamefont {Cederbaum},\ and\ \citenamefont
  {Walter}}]{Schirmer83}%
  \BibitemOpen
  \bibfield  {author} {\bibinfo {author} {\bibfnamefont {J.}~\bibnamefont
  {Schirmer}}, \bibinfo {author} {\bibfnamefont {L.~S.}\ \bibnamefont
  {Cederbaum}}, \ and\ \bibinfo {author} {\bibfnamefont {O.}~\bibnamefont
  {Walter}},\ }\href {\doibase 10.1103/PhysRevA.28.1237} {\bibfield  {journal}
  {\bibinfo  {journal} {Phys. Rev. A}\ }\textbf {\bibinfo {volume} {28}},\
  \bibinfo {pages} {1237} (\bibinfo {year} {1983})}\BibitemShut {NoStop}%
\bibitem [{\citenamefont {Binder}\ \emph
  {et~al.}(2013{\natexlab{a}})\citenamefont {Binder}, \citenamefont
  {Langhammer}, \citenamefont {Calci}, \citenamefont {Navr{\'a}til},\ and\
  \citenamefont {Roth}}]{BiLa13}%
  \BibitemOpen
  \bibfield  {author} {\bibinfo {author} {\bibfnamefont {S.}~\bibnamefont
  {Binder}}, \bibinfo {author} {\bibfnamefont {J.}~\bibnamefont {Langhammer}},
  \bibinfo {author} {\bibfnamefont {A.}~\bibnamefont {Calci}}, \bibinfo
  {author} {\bibfnamefont {P.}~\bibnamefont {Navr{\'a}til}}, \ and\ \bibinfo
  {author} {\bibfnamefont {R.}~\bibnamefont {Roth}},\ }\href@noop {} {\bibfield
   {journal} {\bibinfo  {journal} {Physical Review C}\ }\textbf {\bibinfo
  {volume} {87}},\ \bibinfo {pages} {021303} (\bibinfo {year}
  {2013}{\natexlab{a}})}\BibitemShut {NoStop}%
\bibitem [{\citenamefont {Binder}\ \emph
  {et~al.}(2013{\natexlab{b}})\citenamefont {Binder}, \citenamefont {Piecuch},
  \citenamefont {Calci}, \citenamefont {Langhammer}, \citenamefont
  {Navr{\'a}til},\ and\ \citenamefont {Roth}}]{BiPi13}%
  \BibitemOpen
  \bibfield  {author} {\bibinfo {author} {\bibfnamefont {S.}~\bibnamefont
  {Binder}}, \bibinfo {author} {\bibfnamefont {P.}~\bibnamefont {Piecuch}},
  \bibinfo {author} {\bibfnamefont {A.}~\bibnamefont {Calci}}, \bibinfo
  {author} {\bibfnamefont {J.}~\bibnamefont {Langhammer}}, \bibinfo {author}
  {\bibfnamefont {P.}~\bibnamefont {Navr{\'a}til}}, \ and\ \bibinfo {author}
  {\bibfnamefont {R.}~\bibnamefont {Roth}},\ }\href@noop {} {\bibfield
  {journal} {\bibinfo  {journal} {Physical Review C}\ }\textbf {\bibinfo
  {volume} {88}},\ \bibinfo {pages} {054319} (\bibinfo {year}
  {2013}{\natexlab{b}})}\BibitemShut {NoStop}%
\end{thebibliography}%

\end{document}